\documentclass[graybox]{svmult}

\usepackage{amsmath,amssymb,bm}
\usepackage{mathptmx}       
\usepackage{helvet}         
\usepackage{courier}        
\usepackage{type1cm}        
\usepackage{makeidx}         
\usepackage{graphicx}        
\usepackage{multicol}        
\usepackage[bottom]{footmisc}

\makeindex             

\def\hii{H{\small{II}}\relax}
\def\arcsec{\hbox{$^{\prime\prime}$}}
\def\Q{\ifmmode\mathcal{Q}\else$\mathcal{Q}$\fi}
\newcommand{\lsim}{\ \raise
-2.truept\hbox{\rlap{\hbox{$\sim$}}\raise5.truept\hbox{$<$}\ }}
\newcommand{\gsim}{\ \raise
-2.truept\hbox{\rlap{\hbox{$\sim$}}\raise5.truept\hbox{$>$}\ }}

\begin{document}

\title*{Hierarchically Clustered Star Formation in the Magellanic Clouds}
\titlerunning{Hierarchically Clustered Star Formation in the Magellanic Clouds} 
\author{Dimitrios A. Gouliermis \and Stefan Schmeja  \and 
Volker Ossenkopf \and Ralf S. Klessen \and Andrew E. Dolphin}
\authorrunning{D. A. Gouliermis et al.} 
\institute{Dimitrios A. Gouliermis 
\and Ralf S. Klessen \at Zentrum f\"ur Astronomie der Universit\"at Heidelberg, 
Institut f\"ur Theoretische Astrophysik, Albert-Ueberle-Str.~2, 69120 
Heidelberg, Germany\\ \email{dgoulier@mpia-hd.mpg.de, klessen@uni-heidelberg.de}
\and Stefan Schmeja \at Zentrum f\"{u}r Astronomie der Universit\"{a}t Heidelberg,
Astronomisches Rechen-Institut, M\"{o}nchhofstr. 12-14, 69120 
Heidelberg, Germany \email{sschmeja@ari.uni-heidelberg.de}
\and Volker Ossenkopf \at I. Physikalisches Institut der Universit\"{a}t zu K\"{o}ln, 
Z\"{u}lpicher Stra{\ss}e 77, 50937 K\"{o}ln, Germany \email{ossk@ph1.uni-koeln.de}
\and Andrew E. Dolphin \at Raytheon Company, PO Box 11337, Tucson, AZ 85734, USA
\email{adolphin@raytheon.com}
}
%
%
\maketitle

\abstract*{We present a cluster analysis of the bright main-sequence and faint 
pre--main-sequence stellar populations of a field $\sim 90 \times 90$\,pc centered 
on the \hii\ region NGC~346/N66 in the Small Magellanic Cloud, from imaging with 
HST/ACS. We extend our earlier analysis on the stellar cluster population in the region 
to characterize the structuring behavior of young stars in the region as a whole 
with the use of stellar density maps interpreted through techniques designed for the study 
of the ISM structuring. In particular, we demonstrate with Cartwrigth \& Whitworth's \Q\ 
parameter, dendrograms, and the $\Delta$-variance wavelet transform technique that 
the young stellar populations in the region NGC~346/N66 are highly hierarchically 
clustered, in agreement with other regions in the Magellanic Clouds observed with HST. 
The origin of this hierarchy is currently under investigation.}

\abstract{We present a cluster analysis of the bright main-sequence and faint 
pre--main-sequence stellar populations of a field $\sim 90 \times 90$\,pc centered 
on the \hii\ region NGC~346/N66 in the Small Magellanic Cloud, from imaging with 
HST/ACS. We extend our earlier analysis on the stellar cluster population in the region 
to characterize the structuring behavior of young stars in the region as a whole 
with the use of stellar density maps interpreted through techniques designed for the study 
of the ISM structuring. In particular, we demonstrate with Cartwrigth \& Whitworth's \Q\ 
parameter, dendrograms, and the $\Delta$-variance wavelet transform technique that 
the young stellar populations in the region NGC~346/N66 are highly hierarchically 
clustered, in agreement with other regions in the Magellanic Clouds observed with HST. 
The origin of this hierarchy is currently under investigation.
 }

\section{Method: The identification of stellar clusters}
\label{sec:1}

\begin{figure}[t]
\centerline{\includegraphics[scale=.575]{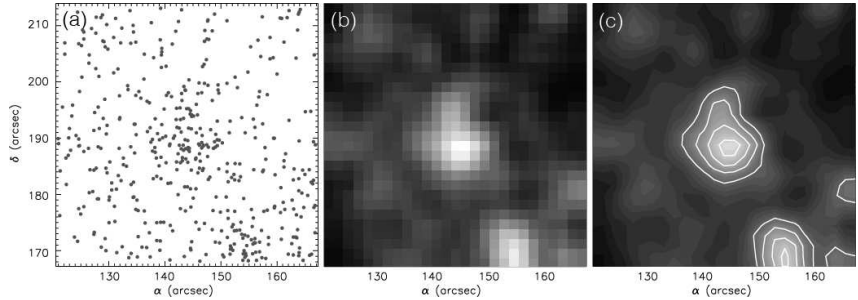}}
\caption{Schematic of the star-count process. (a) The chart of an observed stellar sample. 
(b) The corresponding stellar density 
map, after counting stars in quadrilateral grid of elements (pixels) of size $1.8\arcsec$ each, and after 
filtering the map with a Gaussian of FWHM\,$\simeq2.8$\,px ($\sim$\,5\arcsec). (c) The corresponding 
isodensity contour map. Isopleths at levels \gsim\,3$\sigma$ are indicated with white lines. 
\label{fig:1}}
\end{figure}

For the investigation of the clustering behavior of stars it is necessary to thoroughly 
characterize distinct concentrations of stars, which can only be achieved by the accurate 
identification of individual stellar clusters. Considering the importance of this process, 
different identification methods were developed, which can be classified in two families.
The first, represented by {\sl friend of friend} algorithms and {\sl cluster analysis} techniques, e.g., 
\cite{battinelli96}, are designed for limited samples of observed stars, 
and thus are based on linking individual stars into coherent stellar groups. These methods 
are recently superseded by {\sl minimum spanning trees}, e.g.,  \cite{bastian09}. The second
family of identification codes, represented by {\sl nearest-neighbors} and {\sl star-counts}, make
use of surface stellar density maps constructed from rich observed stellar samples. 
Distinct stellar systems are identified as statistically significant over-densities in respect to the average stellar 
density in the observed regions, e.g., \cite{gouliermis10}. Tests on artificial clusters of various density 
gradients and shapes showed that the latter (density) techniques are more robust in detecting 
real stellar concentrations, provided that rich stellar samples are available \cite{schmeja11}. A 
schematic representation of stellar density maps constructed with star-counts is  shown in 
Fig.~\ref{fig:1}.

\begin{figure}[t]
\sidecaption[t]
\includegraphics[scale=.375]{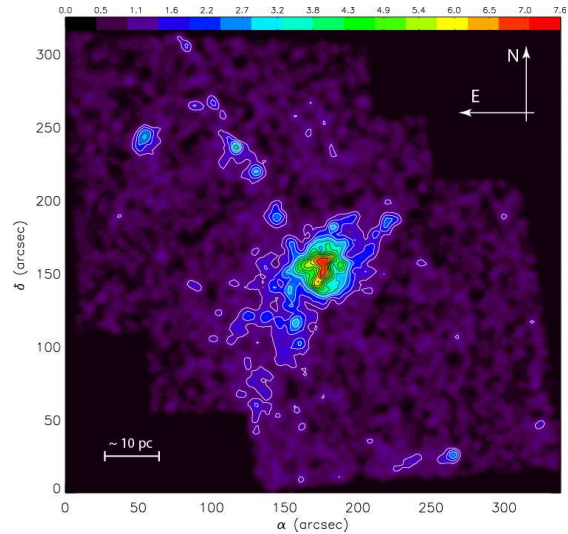}
\caption{Isodensity contour map from star-counts of the 
young bright main-sequence and faint PMS populations identified with HST/ACS in the 
region of NGC~346 in the SMC. Lines represent isopleths of significance \gsim\,$1\sigma$. 
Apart from the dominating central large stellar aggregate, 
there are peripheral young sub-clusters, revealed as statistically important
stellar concentrations. The central aggregate, denoted by the 1$\sigma$ isopleth, encompass 
various distinct sub-groups, which appear at higher density thresholds. NGC~346 itself 
appears at \gsim\,3$\sigma$ significance.
\label{fig:2}}
\end{figure}

\section{Data: Stellar clustering in the region NGC~346/N66}
\label{sec:2}
 
One of the most prominent bright stellar systems in the Small Magellanic Cloud 
(SMC) is the stellar association NGC~346, related to the \hii\ region 
LHA~115-N66 \cite{henize56}, the brightest in this galaxy. This system 
appears in partially-resolved observations form the ground as a single stellar 
concentration, but recent imaging with the {\sl Advanced Camera for Surveys} onboard 
the Hubble Space Telescope (HST) allowed the detection of smaller  sub-clusters 
within the boundaries of the \hii\ nebula. The images were collected within 
the HST GO Program 10248 and were retrieved from the HST Data Archive. 
Their photometry demonstrated that the faint young stellar populations in the 
region are still in their pre--main-sequence (PMS) phase, and revealed a plethora  
of sub-solar PMS stars \cite{gouliermis06}. Our  {\sl nearest-neighbor} cluster 
analysis of the observed young stellar populations, i.e., the bright main-sequence 
(down to $m_{555} \lsim 21$) and the faint PMS stars, revealed a significant number 
of smaller, previously unresolved, young stellar sub-clusters  \cite{schmeja09}. 
This clustering behavior of young stars in NGC~346 is further demonstrated here 
by the stellar density contour map of Fig~\ref{fig:2}, constructed with star-counts.

\section{Results: Hierarchical clustering of young stars}
\label{sec:3}

The map of Fig.~\ref{fig:2} shows significant sub-structure, in particular within the 1$\sigma$ 
boundaries of the central dominant stellar aggregate. This structuring behavior indicates
hierarchy. The minimum spanning tree (MST) of the young stars in the whole region allows to 
determine the statistical \Q\ parameter, introduced by \cite{cw04}. This parameter is a measure 
of the fractal dimension $D$ of a stellar group, permitting to  
distinguish between centrally concentrated clusters and hierarchical clusters with fractal 
substructure. The application of the MST to our data shows that the region NGC~346/N66 is
highly hierarchical with a \Q\ that corresponds to a fractal dimension $D \simeq 2.5$.

\begin{figure}[t]
\sidecaption[t]
\includegraphics[scale=.375]{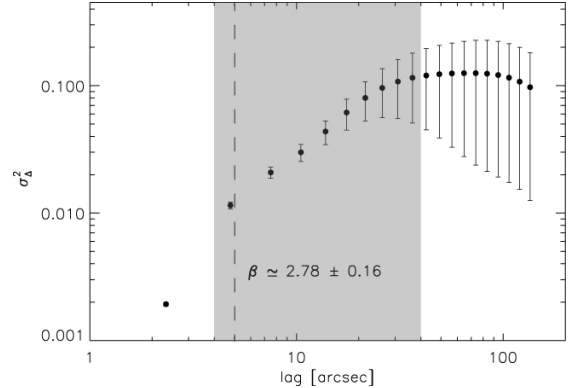}
\caption{The $\Delta$-variance spectrum of the surface stellar density map of the 
entire region of NGC~346/N66. This analysis shows that the young stellar populations in this region are
hierarchically structured up to length-scales of $\sim$\,40\arcsec. The spectral index $\beta$ 
is determined from the fit of the spectrum for data between lags 4\arcsec and 13\arcsec (indicated by the gray 
shaded area). The dashed line provides the used virtual beamsize (5\arcsec). 
\label{fig:3}}
\end{figure}

Constructing surface stellar density maps allows us to further characterize the clustering behavior 
of stars with the application of tools, which are originally designed for the study of the structuring of 
the interstellar medium (ISM), as observed at far-infrared or longer wavelengths. The so-called 
{\sl dendrograms} are used for the visualization of hierarchy through structural trees 
\cite{rosolowsky08}. The dendrogram of the stellar density map of NGC~346 demonstrates that
the observed hierarchy is mostly due to the substructure in the dominant stellar aggregate.
The $\Delta$-variance analysis \cite{stutzki98, ossenkopf08} is a robust structure analysis method
that measures the amount of structure on a given scale $l$. In 
principle the $\Delta$-variance is directly related to the power spectrum of the map, and thus for
a  power law spectrum of index $-\beta$, $\Delta$-variance also follows a power law, $\displaystyle 
\sigma_\Delta^2 \propto l^\alpha$, with $\alpha={\beta-2}$. 
The application of the $\Delta$-variance analysis on the surface stellar density map of NGC~346 
verifies that indeed the clustering of the young stars in the region is self-similar (Fig.~\ref{fig:3}),
with a spectral index $\beta \simeq 2.8$, corresponding to a fractal dimension $D=2.6$ of the 
corresponding fractional Brownian motion structure \cite{stutzki98}, similar to that previously derived for Galactic 
molecular clouds. Self-similarity appears to brake, i.e., we find different hierarchical properties for the 
short-range scaling and the behavior at the overall scale of the region, at length-scales $l \geq 25$\,px, 
corresponding to physical scales of $\sim$\,40\arcsec ($\sim 11$\,pc at the distance of the SMC).



\begin{acknowledgement}
D.A.G., S.S. and V.O. kindly acknowledge support by the German Research Foundation (DFG) through grants GO~1659/3-1, SFB~881 and OS~177/2-1
respectively. Based on observations made with the NASA/ESA {\em Hubble Space Telescope}, obtained from the data archive at the 
Space Telescope Science Institute (STScI). STScI is operated by the Association of Universities for Research in Astronomy, Inc.\ under 
NASA contract NAS 5-26555.
\end{acknowledgement}

\end{document}